\documentclass[nopreprintline,5p,number]{elsarticle}

\usepackage[english]{babel}
\usepackage{microtype}
\usepackage{csquotes}
\usepackage{acro}
\usepackage{amsmath,amssymb,mathtools}
\usepackage{derivative,slashed,braket,tensor}
\usepackage{siunitx}
\usepackage{graphicx}
\usepackage[hidelinks]{hyperref}
\usepackage[capitalise]{cleveref}
\usepackage{geometry}
\newlength{\arxivmargin}
\newlength{\arxivpaperwidth}
\biboptions{sort&compress}

\acsetup{case-insensitive}
\AddToHook{env/abstract/end}{\acresetall}

\DeclareAcronym{qed}{
short={QED},
long={quantum electrodynamics},
plural={},
first-style=short,
}

\DeclareAcronym{qcd}{
short={QCD},
long={quantum chromodynamics},
plural={},
first-style=short,
}

\DeclareAcronym{eft}{
short={EFT},
long={effective field theory},
long-plural-form={effective field theories},
indefinite={an},
}

\DeclareAcronym{nrqcd}{
short={NRQCD},
long={nonrelativistic \acs*{QCD}},
plural={},
}

\DeclareAcronym{pnrqcd}{
short={pNRQCD},
long={potential \acl*{NRQCD}},
plural={},
}

\DeclareAcronym{lo}{
short={LO},
long={leading order},
plural={},
}

\DeclareAcronym{m1}{
short={M1},
long={magnetic dipole},
plural={},
}

\DeclareAcronym{hfs}{
short={HFS},
long={hyperfine splitting},
}

\DeclareAcronym{pdg}{
short={PDG},
long={Particle Data Group},
plural={},
}

\DeclareAcronym{erc}{
short={ERC},
long={European Research Council},
plural={},
}

\AddBabelHook[english]{USenglish}{afterextras}{

}

\DeclareMathOperator{\BF}{\mathcal{B}}
\DeclareMathOperator{\Order}{\mathcal{O}}

\DeclarePairedDelimiter{\abs}{\lvert}{\rvert}
\DeclarePairedDelimiterXPP{\evalat}[2]{}{.}{\rvert}{_{#2}}{#1}

\newcommand*{\given}[1][]{\nonscript\:#1\vert\nonscript\:}

\newcommand*{\alphas}{\alpha_{\mathrm{s}}}
\newcommand*{\Jpsi}{J/\psi}
\newcommand*{\etac}{\eta_{c}}
\newcommand*{\hfs}{\Delta E_{\textnormal{\acs*{HFS}}}}

\newcommand*{\erc}[1]{\href{https://doi.org/10.3030/#1}{#1}}

\makeatletter
\let\orig@pprintTitle\ps@pprintTitle
\def\ps@pprintTitle{%
  \orig@pprintTitle
  \def\@oddhead{\reset@font\hfill\small{TUM-EFT 215/26}}%
  \let\@evenhead\@oddhead}
\makeatother

\begin{document}

\begin{frontmatter}

\title{Extraction of charmonium branching fractions from \texorpdfstring{$\Jpsi\to\gamma\etac$}{J/psi -> gamma etac} radiative decays}

\author[1]{Magnus~C.~Schaaf}
\ead{magnus.schaaf@tum.de}

\author[1]{Antonio~Vairo}
\ead{antonio.vairo@tum.de}

\affiliation[1]{%
organization={Technical University of Munich, TUM School of Natural Sciences, Physics Department},
addressline={James-Franck-Str.~1},
postcode={85748},
city={Garching},
country={Germany},
postcodesep={}
}

\begin{abstract}
We assess the tension between theoretical predictions and the values quoted by the \ac{PDG} for the partial decay width and branching fraction associated with the radiative charmonium decay $\Jpsi\to\gamma\etac$.
A profile scan over the most recent \ac{PDG} data depending on the branching fraction $\BF(\Jpsi\to\gamma\etac)$ suggests that the correlation between measured branching fractions is compatible with lattice~\ac{QCD} determinations of the partial decay widths $\Gamma(\Jpsi\to\gamma\etac)$ and $\Gamma(\etac\to\gamma\gamma)$.
We propose a theoretically grounded photon line shape for the radiative decay spectrum and a prescription for the extraction of (product) branching fractions involving the \ac{M1} transition $\Jpsi\to\gamma\etac$.
This approach obviates the need to modify the photon energy spectrum line shape using empirical damping functions, as done in the most recent experimental extractions of $\BF(\Jpsi\to\gamma\etac)$ from the photon line shape, thereby eliminating an inherent ambiguity in the determination of the derived observables.
\end{abstract}

\end{frontmatter}

\section{Introduction}
Since its discovery in 1974~\cite{E598:1974sol,SLAC-SP-017:1974ind}, the $\Jpsi$ has provided numerous observables for phenomenological studies of the strong interaction in the framework of \ac{QCD}~\cite{QuarkoniumWorkingGroup:2004kpm,Brambilla:2010cs}.
Among these, the \ac{M1} transition $\Jpsi\to\gamma\etac$ is of particular interest.
From a theoretical perspective, the process exhibits a hierarchy of well-separated energy scales, allowing it to be treated within a nonrelativistic expansion in \iac{EFT} framework, such as \ac{pNRQCD}~\cite{Brambilla:2005zw,Brambilla:2010ey,Pineda:2013lta}.
Because it is an \emph{allowed} \ac{M1} transition, the transition rate at \ac{LO} in the nonrelativistic expansion is completely accessible by weakly-coupled perturbation theory.
The transition $\Jpsi\to\gamma\etac$ also provides a clean observable for lattice~\ac{QCD} calculations, as the relevant states lie below the open-charm threshold and the process involves only a single photon emission~\cite{Dudek:2006ej,Dudek:2009kk,Chen:2011kpa,Becirevic:2012dc,Donald:2012ga,Gui:2019dtm,Delaney:2023fsc,Colquhoun:2023zbc,Meng:2024axn}.
Experimentally, the radiative decay offers a distinctive signature with well-defined kinematics and benefits from the copious production of $\Jpsi$ at $e^+e^-$ colliders.

To date, the branching fraction $\BF(\Jpsi\to\gamma\etac)$ has been measured from the photon energy spectrum line shape in the region close to the $\etac$ peak of the radiative decay $\Jpsi\to\gamma X$ by the Crystal~Ball~\cite{Gaiser:1985ix}, CLEO~\cite{CLEO:2008pln}, and KEDR~\cite{Anashin:2010nr,Anashin:2014wva} experiments.
The \ac{PDG} quotes an average value of $\BF_{\textnormal{\acs*{PDG}}}^{\text{(av)}}(\Jpsi\to\gamma\etac)=\qty{1.7(4)}{\percent}$~\cite{ParticleDataGroup:2026mpi}, with the uncertainty scaled by a factor of \num{1.5}, reflecting the observed spread in the measurements~\footnote{
	The \ac*{PDG} average and fit only include the measurements reported by Crystal~Ball~\cite{Gaiser:1985ix} and CLEO~\cite{CLEO:2008pln}, but exclude the one reported by KEDR~\cite{Anashin:2014wva} owing to the absence of systematic uncertainties.
}.
Most of the theoretical determinations suggest values higher than the \ac{PDG} average.
For example, the weakly-coupled \ac{pNRQCD} determination of~\cite{Pineda:2013lta} gives $\Gamma_{\textnormal{\acs*{pNRQCD}}}(\Jpsi\to\gamma\etac)=\qty{2.12(40)}{\keV}$, corresponding to a branching fraction of \qty{2.29}{\percent} (normalized to the total width $\Gamma_{\Jpsi}$~\cite{ParticleDataGroup:2026mpi}), which lies above the experimental average and so do the latest, very precise, lattice~\ac{QCD} determinations~\cite{Becirevic:2012dc,Gui:2019dtm,Colquhoun:2023zbc,Meng:2024axn};
see \cref{fig:scan:th}.

The determination of $\BF(\Jpsi\to\gamma\etac)$ affects the extraction of several $\etac$ branching fractions.
Individual measurements of $\BF(\etac\to X_i)$ are commonly determined in exclusive processes of the type $\Jpsi\to\gamma\etac\to\gamma X_i$ from the product branching fractions $\BF(\Jpsi\to\gamma\etac)\times\BF(\etac\to X_i)$ and thus depend directly on $\BF(\Jpsi\to\gamma\etac)$.
The dominant uncertainty source in these measurements is typically the normalization to $\BF_{\textnormal{\acs*{PDG}}}(\Jpsi\to\gamma\etac)$.
In addition, the $\etac$ total decay width, 10 combinations of partial decay widths, and 38 branching ratios are also determined in a multiparticle fit by the \ac{PDG}, with the $\BF(\Jpsi\to\gamma\etac)$ and $\BF(\Jpsi\to\gamma\etac)\times\BF(\etac\to X_i)$ measurements serving as direct inputs~\footnote{
	Before 2024, the total and partial decay widths were determined in a single particle constrained fit by the \ac*{PDG} and the $\BF(\Jpsi\to\gamma\etac)$ measurements entered indirectly via the normalization of $\BF(\Jpsi\to\gamma\etac)\times\BF(\etac\to X_i)$ to the average $\BF_{\textnormal{\acs*{PDG}}}^{\text{(av)}}(\Jpsi\to\gamma\etac)$ or to $\BF_{\textnormal{Crystal~Ball}}(\Jpsi\to\gamma\etac)$.
}.
From this fit,  based on 115 measurements with 19 parameters, the \ac{PDG} obtains $\BF_{\textnormal{\acs*{PDG}}}^{\text{(fit)}}(\Jpsi\to\gamma\etac)= \qty{1.82(15)}{\percent}$ and $\Gamma_{\textnormal{\acs*{PDG}}}^{\text{(fit)}}(\etac\to\gamma\gamma)= \qty{6.4(5)}{\keV}$~\cite{ParticleDataGroup:2026mpi}.
The width $\Gamma(\etac\to\gamma\gamma)$ can also be calculated on the lattice~\cite{CLQCD:2020njc,Liu:2020qfz,Meng:2021ecs,Colquhoun:2023zbc}.
While recent high-precision lattice~\ac{QCD} determinations~\cite{Meng:2021ecs,Colquhoun:2023zbc} have been in tension with previously reported \ac{PDG} values~\cite{ParticleDataGroup:2024cfk}, the latest \ac{PDG} update~\cite{ParticleDataGroup:2026mpi} largely resolves this discrepancy; see \cref{fig:scan:th}.

The above contrasts point towards an inconsistency in the extraction of $\BF(\Jpsi\to\gamma\etac)$ from data.
In the CLEO and KEDR analyses~\cite{CLEO:2008pln,Anashin:2010nr,Anashin:2014wva}, the branching fraction is determined by fitting a line shape for the photon energy spectrum to measured data of the radiative decay, from which the signal yield is extracted.
The line shape is modeled by a relativistic Breit--Wigner distribution, modified by an $E_\gamma^3$ factor, with $E_\gamma$ denoting the emitted photon's energy.
This factor naturally arises in theoretical calculations of the transition width~\cite{Brambilla:2005zw} and is necessary to reconcile the spectrum with Low's theorem, equivalent to the modification of the Ore--Powell spectrum in \ac{QED}~\cite{Manohar:2003xv}.
Experimentally, it accounts for the observed asymmetry in the photon energy spectrum~\cite{CLEO:2008pln,Brambilla:2010ey}, but it also leads to a growing tail at high energies.
If the $\etac$ yield is computed by just integrating the line shape, the result is (power) divergent and requires the introduction of a cut-off.
Such a cut-off is explicitly introduced in the CLEO~\cite{CLEO:2008pln}, KEDR~\cite{Anashin:2010nr,Anashin:2014wva}, and more recently BESIII extractions~\cite{BESIII:2024rex,BESIII:2025vdn} of $\BF(\Jpsi\to\gamma\etac)$  through the use of \textit{ad hoc} damping functions that suppress the signal contribution far from the $\etac$ peak region of the photon energy spectrum.
Since there is no theoretical justification for any specific form of such damping functions, their introduction leads to unphysical, cut-off-dependent extractions of the branching fraction.
In a different theoretical study~\cite{Segovia:2021bjb}, the power-like tail of the line shape is subtracted in dimensional regularization.

In this Letter, we examine the dependence of the partial decay width $\Gamma(\etac\to\gamma\gamma)$ on the value of the branching fraction $\BF(\Jpsi\to\gamma\etac)$ in a \ac{PDG}-like multiparticle fit.
We find that the fit can yield results compatible with recent lattice~\ac{QCD} calculations for larger values of $\BF(\Jpsi\to\gamma\etac)$, demonstrating that a reliable extraction of $\BF(\Jpsi\to\gamma\etac)$ from measured data is essential.
We revisit the theoretical description of the \ac{M1} transition and propose a different method to extract the $\Jpsi\to\gamma\etac$ signal yield from the photon energy spectrum line shape.
The method is cut-off independent and therefore eliminates the ambiguity affecting present determinations of $\BF(\Jpsi\to\gamma\etac)$ from the line shape.
To assess the impact on the measured branching fractions, we reevaluate experimental data on the radiative decay with the proposed line shape.

\section{Multiparticle fit}
\begin{figure}[t]
\centering
\includegraphics{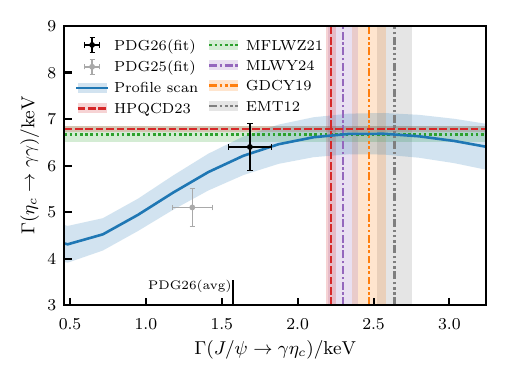}
\caption{%
Fit values of $\Gamma(\etac\to\gamma\gamma)$ as a function of $\BF(\Jpsi\to\gamma\etac)$;
the latter is multiplied by the $\Jpsi$ total width $\Gamma_{\Jpsi}$ quoted in Ref.~\cite{ParticleDataGroup:2026mpi} to obtain the partial decay width.
The black and gray dots with error bars show the results from the 2026~\cite{ParticleDataGroup:2026mpi} and 2025~\cite{ParticleDataGroup:2024cfk} \ac{PDG} multiparticle fit, respectively;
the black marker indicates the \ac{PDG} average value of $\Gamma(\Jpsi\to\gamma\etac)$~\cite{ParticleDataGroup:2026mpi}.
The blue solid curve shows the profile scan, with the shaded band indicating the uncertainty.
The horizontal and vertical lines denote values of $\Gamma(\etac\to\gamma\gamma)$ and $\Gamma(\Jpsi\to\gamma\etac)$ from lattice~\ac{QCD} calculations, respectively, with shaded bands indicating their uncertainties.
The reference lines are red dashed (HPQCD23)~\cite{Colquhoun:2023zbc}, green dotted (MFLWZ21)~\cite{Meng:2021ecs}, purple dash dotted (MLWY24)~\cite{Meng:2024axn}, orange dash double dotted (GDCY19)~\cite{Gui:2019dtm}, and gray dash triple dotted (EMT12)~\cite{Becirevic:2012dc}.
}\label{fig:scan:th}
\end{figure}

To obtain the dependence of $\Gamma(\etac\to\gamma\gamma)$ on $\BF(\Jpsi\to\gamma\etac)$ we perform a multiparticle fit following the \ac{PDG}~\cite{ParticleDataGroup:2026mpi}, but fix the branching fraction $\BF(\Jpsi\to\gamma\etac)$ to values from \qtyrange{0.5}{3.5}{\percent} while omitting the respective Crystal~Ball and CLEO measurements~\cite{Gaiser:1985ix,CLEO:2008pln}.
The resulting fitted values for $\Gamma(\etac\to\gamma\gamma)$ are displayed in \cref{fig:scan:th}, together with results from recent lattice~\ac{QCD} calculations~\footnote{
    We attribute the small difference between our scan and the \ac{PDG} value to the independent implementation of the fit procedure, including differences in the minimization algorithm and convergence criteria.
    The small localized non-smooth structure near the small-width end of the profile curve is a consequence of the graphic interpolation over a finite set of scan points.
}.
While the lattice predictions are in tension with the latest \ac{PDG} fit, the profile scan shows that the measured data could in principle be compatible with recent lattice~\ac{QCD} determinations.
In fact, the profile scan crosses the HPQCD predictions for $\Gamma(\etac\to\gamma\gamma)$ and $\Gamma(\Jpsi\to\gamma\etac)$~\cite{Colquhoun:2023zbc}, which is not guaranteed \textit{a priori}.
The fit quality is sensitive to the inclusion of two measurements newly added in the 2026 \ac{PDG} update~\cite{ParticleDataGroup:2026mpi} (a total width~\cite{BESIII:2024rdn} and a product branching fraction~\cite{BESIII:2024rex} measurement), which resolves part of the tension with lattice~\ac{QCD} determinations of $\Gamma(\etac\to\gamma\gamma)$, but also results in a worse overall fit quality ($\chi^2=\num{215.4}$ for $\num{96}$ degrees of freedom).
This may reflect a systematic issue in the current $\Jpsi\to\gamma\etac$ data and motivates a reassessment of the related measurements.

\section{Line shape}
As previously recognized~\cite{Brambilla:2010ey}, the reported values for the branching fraction corresponding to the $\Jpsi\to\gamma\etac$ transition in Refs.~\cite{CLEO:2008pln,Anashin:2010nr} actually refer to the process $\Jpsi\to\gamma\etac\to\gamma X$, where the intermediate state can be off-shell, and $X$ denotes its final decay modes.
The problem lies in the fact that the function that parametrizes the differential $\Jpsi\to\gamma\etac\to\gamma X$ width, which is a Breit--Wigner distribution times $E_\gamma^3$, contains both the contribution from the \ac{M1} transition $\Jpsi\to\gamma\etac$ around the $\etac$ peak and the Ore--Powell spectrum at high energies~\cite{Manohar:2003xv}.

More specifically, for photon energies $E_\gamma$ close to the $1S$ \ac{HFS} $\hfs \equiv  M_{\Jpsi}-M_{\etac}$, the differential decay width for the inclusive decay $\Jpsi\to\gamma X$ can be obtained within \ac{pNRQCD} as~\cite{Manohar:2003xv,Brambilla:2010ey}
\begin{equation}\label{eq:differential_decay}
\begin{split}
	\odv{\Gamma_{\Jpsi\to\gamma X}}{E_\gamma}
    \mspace{16mu}&\underset{\crampedclap{E_\gamma \lesssim \qty{500}{\MeV}}}{\approx}\mspace{16mu}
	\odv{\Gamma_{\Jpsi\to\gamma\etac\to\gamma X}^{\textnormal{(\acs*{M1})}}}{E_\gamma}
    \\
	&\underset{\crampedclap{\textnormal{\acs*{LO}}}}{=}\mspace{3mu}
	A\frac{E_\gamma^3}{\pi}\frac{\Gamma_{\etac}/2}{(\hfs-E_\gamma)^2 + \Gamma_{\etac}^2/4}
	\,,
\end{split}
\end{equation}
where $\Gamma_{\etac}$ represents the total width of the $\etac$, in the region of interest $E_\gamma \lesssim \qty{500}{\MeV}$, which is about the $\Jpsi$ binding energy~\cite{Brambilla:2010ey}.
Here and in the following, \acs*{LO} refers to the leading order in the nonrelativistic expansion, for which
\begin{equation}
	A
	=
	\frac{4}{3}\frac{{c_{\mathrm{F}}^{\textnormal{em}}}^2 \alpha e_c^2}{m_c^2}
	\,,
\end{equation}
where $\alpha$ is the fine structure constant, $e_c=2/3$ is the fraction of electric charge carried by the charm quark, $m_c \approx M_{\Jpsi}/2$ is the charm quark mass, and $c_{\mathrm{F}}^{\textnormal{em}}=1+\kappa_c^{\textnormal{em}}$, with $\kappa_c^{\textnormal{em}}=2\alphas/(3\pi)+\Order(\alphas^2)$ the anomalous magnetic moment of the heavy quark~\cite{Brambilla:2005zw}.
The differential decay width is the product of $E_\gamma^3$ with a nonrelativistic Breit--Wigner distribution.
It vanishes as $\Order(E_\gamma^3)$ for $E_\gamma\to0$ in accordance with Low's theorem and develops an $\Order(E_\gamma)$ Ore--Powell tail in the region $E_\gamma\gg\hfs$.
The latter includes high-energy events not associated with the production of an intermediate $\etac$~\cite{Manohar:2003xv}.
The partial decay width of $\Jpsi\to\gamma\etac$ is recovered from \cref{eq:differential_decay} by taking the on-shell limit $\Gamma_{\etac}\to0$,
\begin{equation}
\begin{split}
	\odv{\Gamma_{\Jpsi\to\gamma\etac}}{E_\gamma}
	&=
	\lim_{\Gamma_{\etac}\to0}\odv{\Gamma_{\Jpsi\to\gamma\etac\to\gamma X}^{\textnormal{(\acs*{M1})}}}{E_\gamma}
	\\
	&\underset{\crampedclap{\textnormal{\acs*{LO}}}}{=}\mspace{3mu}
	A E_\gamma^3 \delta(\hfs-E_\gamma)
	\,,
\end{split}
\end{equation}
where the $\delta$-function fixes the energy of the emitted photon, yielding
\begin{equation}\label{eq:partial_decay_width}
	\Gamma_{\Jpsi\to\gamma\etac}
	\underset{\textnormal{\acs*{LO}}}{=}
	A {(\hfs)}^3
	\,,
\end{equation}
in agreement with well known results~\cite{QuarkoniumWorkingGroup:2004kpm,Brambilla:2005zw}.

The photon energy spectrum line shape near the $1S$ \ac{HFS} is given by
\begin{equation}\label{eq:photon_spectrum}
	\odv{N_\gamma}{E_\gamma}
	=
	N_{\Jpsi}\odv{\BF(\Jpsi\to\gamma X)}{E_\gamma}
	=
	\frac{N_{\Jpsi}}{\Gamma_{\Jpsi}}\odv{\Gamma_{\Jpsi\to\gamma X}}{E_\gamma}
	\,,
\end{equation}
where $N_{\Jpsi}$ and $\Gamma_{\Jpsi}$ are the total number and total width of $\Jpsi$.
Integrating \cref{eq:photon_spectrum} gives the total number of emitted photons $N_\gamma$.
Using \cref{eq:differential_decay} yields a divergent result because the integration domain at high energies violates the conditions under which \cref{eq:differential_decay} is valid.
We are, however, not interested in $N_\gamma$, but in the total number of $\Jpsi\to\gamma\etac$ transitions, $N_{\etac}$, which is given by
\begin{equation}
	N_{\etac}
	=
	N_{\Jpsi}\BF(\Jpsi\to\gamma\etac)
	\,.
\end{equation}
The branching fraction for the \ac{M1} transition can be recovered from the differential decay width $\odv{\Gamma_{\Jpsi\to\gamma X}}/{E_\gamma}$ via
\begin{equation}\label{eq:branching_fraction}
	\BF(\Jpsi\to\gamma\etac)
    =
	\int_0^\infty\frac{\odif{E_\gamma}}{\Gamma_{\Jpsi}}\odv{\Gamma_{\Jpsi\to\gamma X}}{E_\gamma}w(E_\gamma)
	\,,
\end{equation}
where choosing
\begin{equation}
	w(E_\gamma)
	\underset{\textnormal{\acs*{LO}}}{=}
	\pi\frac{\Gamma_{\etac}}{2}\delta(\hfs-E_\gamma)
\end{equation}
reproduces the branching fraction that follows from~\labelcref{eq:partial_decay_width}, when replacing the differential decay width with \cref{eq:differential_decay}.
This is because at \ac{LO} the combination
\[
	\odv{\Gamma_{\Jpsi\to\gamma X}}{E_\gamma} \times \pi\frac{\Gamma_{\etac}}{2}\delta(\hfs-E_\gamma)
\]
coincides with the on-shell limit
\[
	\lim_{\Gamma_{\etac}\to0}\odv{\Gamma_{\Jpsi\to\gamma\etac\to\gamma X}^{\textnormal{(\acs*{M1})}}}{E_\gamma}
	\,.
\]
Hence, $N_{\etac}$ can be obtained from the photon energy spectrum line shape as
\begin{equation}\label{eq:signal_yield}
\begin{split}
	N_{\etac}
	&=
	N_{\Jpsi}\int_0^\infty\frac{\odif{E_\gamma}}{\Gamma_{\Jpsi}}\odv{\Gamma_{\Jpsi\to\gamma X}}{E_\gamma}w(E_\gamma)
	\\
	&=
	\int_0^\infty\odif{E_\gamma}\odv{N_\gamma}{E_\gamma}w(E_\gamma)
	\\
	&\underset{\textnormal{\acs*{LO}}}{=}
	\pi\frac{\Gamma_{\etac}}{2}\evalat*{\odv{N_\gamma}{E_\gamma}}{E_\gamma=\hfs}
	\,.
\end{split}
\end{equation}
This equation provides the number of $\Jpsi\to\gamma\etac$ transitions as a function of parameters entering the line shape~\labelcref{eq:photon_spectrum} only;
it does not depend on a cut-off since it does not involve integrating over the photon energy, and it gives a branching fraction consistent with perturbation theory at \ac{LO}.

After inserting the differential decay width~\labelcref{eq:differential_decay} into \cref{eq:photon_spectrum}, the photon energy spectrum line shape reads
\begin{equation}\label{eq:line_shape}
	\odv{N_\gamma}{E_\gamma}
	\underset{\textnormal{\acs*{LO}}}{=}
	N E_\gamma^3 \frac{\Gamma_{\etac}/2}{(\hfs-E_\gamma)^2 + \Gamma_{\etac}^2/4}
	\,.
\end{equation}
It depends upon three parameters, $\hfs$, $\Gamma_{\etac}$, and an overall normalization $N$ that determines the strength of the signal.
Once they have been fitted to data, we obtain the number of signal events from~\cref{eq:signal_yield}, which is the key ingredient to the measured branching fraction.
In actual fits to experimental data, \cref{eq:line_shape} must be multiplied by an energy-dependent efficiency $\epsilon(E)$ and convolved with a detector resolution function $D(E_\gamma \given E)$,
\begin{equation}\label{eq:line_shape:observed}
	\odv{N_\gamma^{\textnormal{(obs)}}(E_\gamma)}{E_\gamma}
	=
	\int\odif{E}\odv{N_\gamma(E)}{E} \epsilon(E) D(E_\gamma \given E)
	\,,
\end{equation}
to describe the observed events.
The proposed line shape can also be applied to exclusive decays $\etac\to X_i$.
In this case, experiments typically measure the invariant mass $M_{X_i}$ distribution of the $\Jpsi\to\gamma\etac\to\gamma X_i$ decay.
This distribution can then be related to the photon energy spectrum line shape by momentum conservation,
\begin{equation}\label{eq:momentum_conservation}
	E_{\gamma}(M_{X_i})
	=
	\frac{M_{\Jpsi}^2-M_{X_i}^2}{2M_{\Jpsi}}
	\,.
\end{equation}

\Cref{eq:signal_yield,eq:line_shape}, which we propose to extract the number of $\Jpsi\to\gamma\etac$ transitions from the photon energy spectrum line shape, should be contrasted with the ones used in experiments.
The CLEO collaboration uses in its analysis $w(E_\gamma) = e^{-E_\gamma^2/(8\beta^2)}$, with $\beta=\qty{65.0(25)}{\MeV}$~\cite{CLEO:2008pln}.
The KEDR collaboration uses $w(E_\gamma) = \hfs^2/(E_\gamma \hfs + (E_\gamma-\hfs)^2)$ in Ref.~\cite{Anashin:2010nr}, and a function that is unity for $\abs{E_\gamma-\hfs}\le2\Gamma_{\etac}$, vanishes for $\abs{E_\gamma-\hfs}\ge4\Gamma_{\etac}$, and interpolates linearly in between in Ref.~\cite{Anashin:2014wva}.
The BESIII collaboration adopts the CLEO and the first KEDR damping functions in many analyses involving the \ac{M1} transition.
Their extractions rely on the damping functions. 
The ensuing results fail to reproduce the \ac{LO} perturbative result once inserted in \cref{eq:branching_fraction}.
In contrast, \cref{eq:signal_yield} is model independent and consistent with perturbation theory as long as the differential width has the form~\labelcref{eq:differential_decay}.
Another difference between \cref{eq:line_shape} and the parameterizations of the photon energy spectrum used in experimental analyses is that \cref{eq:line_shape} incorporates a nonrelativistic Breit--Wigner distribution, consistently with a nonrelativistic \ac{EFT} approach, while the latter incorporate a relativistic Breit--Wigner distribution.
The difference is of higher order in the nonrelativistic expansion.
The additional terms resummed by the relativistic Breit--Wigner distribution include relativistic corrections that depend on the photon energy and the other parameters of the line shape (i.e.~hyperfine splitting and width)~\cite{Brambilla:2010ey}.
Relativistic corrections that do not depend on the photon energy and the other parameters of the line shape are automatically accounted for to all orders by the theoretical expression~\labelcref{eq:differential_decay}, and incorporated in the overall multiplicative constant.
In the line shape, the overall normalization factor $N$ is then fitted to the measured photon energy spectrum.
The difference between fits with a nonrelativistic and relativistic Breit--Wigner distribution may serve as an estimate of unknown higher-order relativistic corrections.

\section{Branching fractions}
To test the proposed line shape and signal extraction, we apply it alongside existing analyses on $\Jpsi\to\gamma\etac$ radiative decays to determine the respective (product) branching fractions.
Three different observables are examined below.

First, we reevaluate the CLEO determination~\cite{CLEO:2008pln} of the branching fraction $\BF^{\textnormal{(\acs*{M1})}}\equiv\BF(\Jpsi\to\gamma\etac)$ that uses a sum of exclusive $\Jpsi\to\gamma\etac\to\gamma X_i$ decays.
We find
\begin{equation}\label{eq:bf:cleo}
	\BF_{\textnormal{CLEO}}^{\textnormal{(\acs*{M1})}}
	=
	\qty{2.01(13)(27)}{\percent}
	\,.
\end{equation}
This value is almost identical to the original determination $\BF_{\textnormal{CLEO\cite{CLEO:2008pln}}}^{\textnormal{(\acs*{M1})}}=\qty{1.99(9)(31)}{\percent}$ rescaled to the current best value of $\BF_{\textnormal{\acs*{PDG}}}(\psi(2S)\to\pi^+\pi^- \Jpsi)$~\cite{ParticleDataGroup:2026mpi}, despite the fact that \labelcref{eq:bf:cleo} does not depend on an \textit{ad hoc} damping function.

Next, we reevaluate the BESIII determination~\cite{BESIII:2024rex} of the product branching fraction $\BF^{\textnormal{(\acs*{M1},$\gamma\gamma$)}}\equiv\BF(\Jpsi\to\gamma\etac)\times\BF(\etac\to\gamma\gamma)$ from exclusive $\Jpsi\to\gamma\etac\to\gamma\gamma\gamma$ decays.
We find
\begin{equation}
	\BF_{\textnormal{BESIII}}^{\textnormal{(\acs*{M1},$\gamma\gamma$)}}
	=
	\num{5.28(30)(54)e-6}
	\,.
\end{equation}
The result is slightly larger than the product branching fraction reported by BESIII using the CLEO and KEDR damping functions, $\BF_{\textnormal{BESIII\cite{BESIII:2024rex}}}^{\textnormal{(\acs*{M1},$\gamma\gamma$)}}=\num{5.23(26)(30)e-6}$.

Finally, we reevaluate the BESIII determination~\cite{BESIII:2025vdn} of the product branching fraction $\BF^{\textnormal{(\acs*{M1},$p\bar{p}$)}}\equiv\BF(\Jpsi\to\gamma\etac)\times\BF(\etac\to p\bar{p})$ from exclusive $\Jpsi\to\gamma\etac\to\gamma p\bar{p}$ decays.
We find
\begin{equation}\label{eq:bf:bes3:M1:pp}
	\BF_{\textnormal{BESIII}}^{\textnormal{(\acs*{M1},$p\bar{p}$)}}
	=
	\num{2.48(5)(8)e-5}
	\,.
\end{equation}
This result is a significant increase of the reported product branching fraction $\BF_{\textnormal{BESIII\cite{BESIII:2025vdn}}}^{\textnormal{(\acs*{M1},$p\bar{p}$)}}=\num{2.11(2)(7)e-5}$ obtained using the CLEO and KEDR damping functions.

Following the prescription described in Ref.~\cite{BESIII:2025vdn} and combining the product branching fractions $\BF_{\textnormal{BESIII}}^{\textnormal{(\acs*{M1},$\gamma\gamma$)}}$ and $\BF_{\textnormal{BESIII}}^{\textnormal{(\acs*{M1},$p\bar{p}$)}}$ with $\BF^{\textnormal{($\gamma\gamma$,$p\bar{p}$)}}\equiv\BF(\etac\to\gamma\gamma)\times\BF(\etac\to p\bar{p})$ yields
\begin{equation}\label{eq:bfs:bes3:M1}
\begin{split}
	\BF_{\textnormal{BESIII}}^{\textnormal{(\acs*{M1})}}
	&=
	\sqrt{\frac{\BF_{\textnormal{BESIII}}^{\textnormal{(\acs*{M1},$\gamma\gamma$)}}\times\BF_{\textnormal{BESIII}}^{\textnormal{(\acs*{M1},$p\bar{p}$)}}}{\BF_{\textnormal{\cite{BESIII:2025vdn}}}^{\textnormal{($\gamma\gamma$,$p\bar{p}$)}}}}
	\\
	&=
	\qty{2.50(7)(22)}{\percent}
\end{split}
\end{equation}
and
\begin{equation}\label{eq:bfs:bes3:gg}
\begin{split}
	\BF_{\textnormal{BESIII}}^{\textnormal{($\gamma\gamma$)}}
	&=
	\sqrt{\frac{\BF_{\textnormal{BESIII}}^{\textnormal{(\acs*{M1},$\gamma\gamma$)}}\times\BF_{\textnormal{\cite{BESIII:2025vdn}}}^{\textnormal{($\gamma\gamma$,$p\bar{p}$)}}}{\BF_{\textnormal{BESIII}}^{\textnormal{(\acs*{M1},$p\bar{p}$)}}}}
	\\
	&=
	\num{2.11(6)(19)e-4}
	\,,
\end{split}
\end{equation}
where we used $\BF_{\textnormal{\cite{BESIII:2025vdn}}}^{\textnormal{($\gamma\gamma$,$p\bar{p}$)}}=\num{2.1(3)e-7}$~\footnote{
	The average $\BF_{\textnormal{\cite{BESIII:2025vdn}}}^{\textnormal{($\gamma\gamma$,$p\bar{p}$)}}$ represents an updated value of the current \ac*{PDG} average~\cite{ParticleDataGroup:2026mpi}.
}.
Since our extraction of $\BF_{\textnormal{BESIII}}^{\textnormal{(\acs*{M1},$\gamma\gamma$)}}$ is comparable to the BESIII value~\cite{BESIII:2024rex}, the net effect of our determination \labelcref{eq:bf:bes3:M1:pp} is an increase of $\BF_{\textnormal{BESIII}}^{\textnormal{(\acs*{M1})}}$ and a decrease of $\BF_{\textnormal{BESIII}}^{\textnormal{($\gamma\gamma$)}}$ compared to Ref.~\cite{BESIII:2025vdn}.

\section{Discussion and conclusions}
Significant discrepancies between the experimental average and fit values provided by the \ac{PDG}~\cite{ParticleDataGroup:2026mpi} and theoretical predictions of the $\Jpsi\to\gamma\etac$ width can be explained by an inconsistency in the current extraction of the $\Jpsi\to\gamma\etac$ transition width from data.
The \ac{PDG} average is based on the two measurements of Crystal~Ball~\cite{Gaiser:1985ix} and CLEO~\cite{CLEO:2008pln}.
It is pulled down by the former.
In the multiparticle fit by the \ac{PDG}, the results for $\Gamma(\etac\to\gamma\gamma)$ and $\BF(\Jpsi\to\gamma\etac)$ are strongly correlated, as shown by the profile scan in \cref{fig:scan:th}, and a larger value of $\BF(\Jpsi\to\gamma\etac)$ would render \emph{both} $\Gamma(\etac\to\gamma\gamma)$ and $\Gamma(\Jpsi\to\gamma\etac)$ consistent with lattice~\ac{QCD} determinations of these quantities~\cite{Becirevic:2012dc,Gui:2019dtm,Meng:2021ecs,Colquhoun:2023zbc,Meng:2024axn}.

In view of this observation, we have revised the determinations of $\BF(\Jpsi\to\gamma\etac)$ from CLEO data~\cite{CLEO:2008pln}, $\BF(\Jpsi\to\gamma\etac)\times\BF(\etac\to\gamma\gamma)$ from BESIII data~\cite{BESIII:2024rex}, and $\BF(\Jpsi\to\gamma\etac)\times\BF(\etac\to p\bar{p})$ from BESIII data~\cite{BESIII:2025vdn} by proposing a new way to extract the number of produced $\etac$ from the signal.
So far, this number was obtained by convolving the signal with an arbitrary damping function.
In such an approach, however, the result depends on the chosen function and is inconsistent with perturbation theory.
We propose instead to extract the yield of $\etac$ from \cref{eq:signal_yield}, which is the main theoretical finding of this Letter.
Following this approach, there is no need to introduce arbitrary damping functions, and the result is consistent with \ac{LO} perturbation theory.

We obtain from the CLEO data a value of $\BF(\Jpsi\to\gamma\etac)$, given in \cref{eq:bf:cleo}, that is in line with the value quoted by the CLEO collaboration~\cite{CLEO:2008pln}, and from the BESIII data a somewhat larger value, given in \cref{eq:bfs:bes3:M1}, than the one reported by the BESIII collaboration~\cite{BESIII:2025vdn}.
Both values are larger than the fit value obtained by the \ac{PDG} but lead to comparable values of $\BF(\etac\to\gamma\gamma)$ extracted from the combined branching fractions of several exclusive processes of the type $\Jpsi\to\gamma\etac\to\gamma X_i$ entering the fit.

\begin{figure}[t]
\centering
\includegraphics{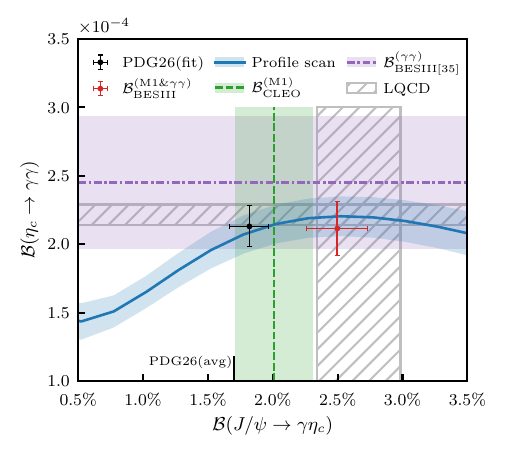}
\caption{%
Fit values of $\BF(\etac\to\gamma\gamma)$ as a function of $\BF(\Jpsi\to\gamma\etac)$.
The black dot with error bars shows the results from the 2026 \ac{PDG} multiparticle fit~\cite{ParticleDataGroup:2026mpi};
the black marker indicates the \ac{PDG} average value of the $\Jpsi\to\gamma\etac$ branching fraction~\cite{ParticleDataGroup:2026mpi}.
The red dot with error bars shows the values from \labelcref{eq:bfs:bes3:M1,eq:bfs:bes3:gg}.
The blue solid curve shows the profile scan, with the shaded band indicating the uncertainty.
The green dashed vertical line shows the values from \cref{eq:bf:cleo} and the purple dash dotted horizontal line shows a recent direct measurement~\cite{BESIII:2026pff} of $\BF(\etac\to\gamma\gamma)$, not depending on the normalization to $\BF(\Jpsi\to\gamma\etac)$;
shaded bands indicate the uncertainties of the corresponding determinations.
The gray hatched regions represent the range of the results of the lattice~\ac{QCD} calculations~\cite{Colquhoun:2023zbc,Meng:2021ecs,Meng:2024axn,Gui:2019dtm,Becirevic:2012dc}, normalized to the \ac{PDG} average of the total widths $\Gamma_{\etac}$ and $\Gamma_{\Jpsi}$~\cite{ParticleDataGroup:2026mpi}.
}\label{fig:scan:ex}
\end{figure}
\Cref{fig:scan:ex} summarizes our findings.
The central value of the green band, which corresponds to the extraction of $\BF(\Jpsi\to\gamma\etac)$ from the CLEO data given in \cref{eq:bf:cleo}, is larger than the \ac{PDG} fit (black dot) and closer to the lattice data, collectively represented by the gray hatched regions.
If we reevaluate the scan of the \ac{PDG} multiparticle fit at this value of the \ac{M1} branching fraction, we find
\begin{equation}
	\BF_{\textnormal{CLEO}}^{\textnormal{(fit)}}(\etac\to\gamma\gamma)
	=
	\num{2.14(14)e-4}
    \,,
\end{equation}
which corresponds to the intersection of the blue solid curve (profile scan) with the green dashed line.
The combined determination of $\BF(\Jpsi\to\gamma\etac)$ from \cref{eq:bfs:bes3:M1} and $\BF(\etac\to\gamma\gamma)$ from \cref{eq:bfs:bes3:gg} based on the most recent BESIII data is represented by the red dot.
Accounting for the uncertainties, it falls in the region preferred by lattice~\ac{QCD} data.
If we reevaluate the scan of the \ac{PDG} multiparticle fit at the BESIII value of the \ac{M1} branching fraction given in \cref{eq:bfs:bes3:M1}, we find
\begin{equation}
	\BF_{\textnormal{BESIII}}^{\textnormal{(fit)}}(\etac\to\gamma\gamma)
	=
	\num{2.20(14)e-4}
    \,.
\end{equation}
Within the uncertainties, this value is compatible with the related extraction~\labelcref{eq:bfs:bes3:gg} and the recent direct measurement~\cite{BESIII:2026pff} of $\BF(\etac\to\gamma\gamma)$.
The differences correspond to the vertical distance of the blue solid curve (profile scan) from the red dot and the  purple dash dotted line, respectively.
It falls inside the region preferred by lattice~\ac{QCD} data.

We conclude that it is possible to extract $\BF(\Jpsi\to\gamma\etac)$ in a model-independent manner and that this extraction provides values of $\BF(\Jpsi\to\gamma\etac)$ and $\BF(\etac\to\gamma\gamma)$ that are consistent with lattice~\ac{QCD} and perturbation theory.
This appears to exclude the need for more exotic theoretical explanations.

\section*{Acknowledgments}
We thank Ryan Mitchell for providing the bin contents underlying Fig.~1 in Ref.~\cite{CLEO:2008pln};
Chang\nobreakdash-Zheng Yuan, Zhijun Li for providing the bin contents underlying Fig.~2 in Ref.~\cite{BESIII:2024rex} as well as the background components and efficiency curve underlying Fig.~2\,(d) and Fig.~4\,(a) in Ref.~\cite{Li:2025pri};
and Chang\nobreakdash-Zheng Yuan, Yipu Liao, Yijia Zeng for providing the bin contents underlying Fig.~1\,(a) in Ref.~\cite{BESIII:2025vdn} and for helpful comments.
M.C.S.\ thanks Svenja Diekmann for valuable discussions.
We thank Sergio~Navas for bringing the correct account of the Crystal~Ball measurement to our attention.
Data published by the \ac{PDG} was accessed via the PDG~API~\cite{Beringer:2024ady,Beringer:2024eae}.
All fits were performed using \textsc{iminuit}/\textsc{Minuit}~\cite{iminuit,James:1975dr}.
The work of M.C.S.\ is supported by the \acs*{ERC} grant EFT-XYZ, \erc{101141922}.

\appendix
\section{Extraction of branching fractions}
The following provides details on the fits performed to obtain the results presented in the main text.
For each observable, we follow the analysis as described in the respective original publications but apply our prescription for the extraction of (product) branching fractions as defined in the main text.

\subsection{\texorpdfstring{$\Jpsi\to\gamma\etac$}{J/psi -> gamma etac}}
\begin{figure}[h]
\centering
\includegraphics{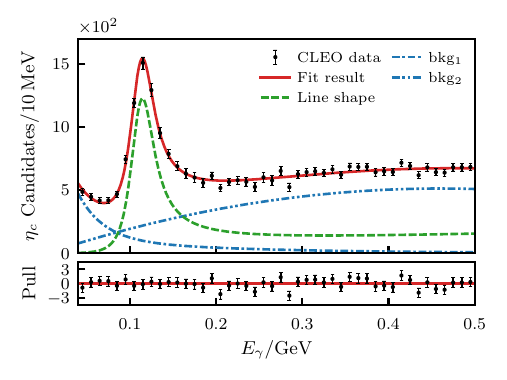}
\caption{%
Fit to the photon energy ($E_\gamma$) spectrum from the sum of exclusive $\Jpsi\to\gamma\etac\to\gamma X_i$ decays.
The black dots with error bars are data points measured by the CLEO experiment~\cite{CLEO:2008pln}.
The red solid curve shows the fit result, with the green dashed curve representing the line shape as given by \cref{eq:line_shape,eq:line_shape:observed}, and the blue dash-dotted and dash-double-dotted curves representing the background components described in Refs.~\cite{CLEO:2008pln,Brambilla:2010ey}.
The lower panel shows the pulls $(N_{\textnormal{data}}-N_{\textnormal{fit}})/\sigma_{\textnormal{data}}$, where $\sigma_{\textnormal{data}}$ denotes the measurement uncertainty.
}\label{fig:cleo}
\end{figure}

We perform a least squares fit following the procedure outlined in Ref.~\cite{CLEO:2008pln} to the photon energy spectrum based on the data points presented in Fig.~1 of Ref.~\cite{CLEO:2008pln}, where the line shape is described by \cref{eq:line_shape,eq:line_shape:observed}, and $D(E_\gamma \given E)$ is modeled by a Gaussian resolution function with a resolution of \qty{4.8}{\MeV}~\cite{CLEO:2008pln}.
The background components $\textnormal{bkg}_{1/2}$ are described in Refs.~\cite{CLEO:2008pln,Brambilla:2010ey} and fitted simultaneously.
The best fit is shown in \cref{fig:cleo}. 
The best-fit parameters are $\hfs=\qty{110.22(66)}{\MeV}$ and $\Gamma_{\etac}=\qty{28.1(19)}{\MeV}$.
The photon energy spectrum line shape develops the anticipated Ore--Powell tail since it describes the sum of exclusive decays $\Jpsi\to\gamma\etac\to\gamma X_i$.
Accordingly, the background component $\textnormal{bkg}_{2}$ is less prominent compared to CLEO's, as it originally describes \enquote{nonsignal $\Jpsi\to\gamma X_i$} events~\cite{CLEO:2008pln}, parts of which are now contained in the new line shape.

We compute the observed signal yield from the fitted line shape using \cref{eq:signal_yield} and the fitted values for $\hfs$, $\Gamma_{\etac}$, and $N$, obtaining $N_{\etac}=\num{5674(324)}$.
The branching fraction follows from
\begin{equation}
	\BF_{\textnormal{CLEO}}^{\textnormal{(\acs*{M1})}}
	=
	\frac{(N_{2S}^{\textnormal{INC}}/N_{2S}^{\textnormal{EXC}})N_{1S}^{\textnormal{EXC}}}{\varepsilon_{2S}^{\textnormal{INC}}(\varepsilon_{1S}^{\textnormal{EXC}}/\varepsilon_{2S}^{\textnormal{EXC}})N_{\psi(2S)}\BF_{\pi\pi}}
	\,,
\end{equation}
where all values except $N_{1S}^{\textnormal{EXC}} \equiv N_{\etac}$ and $\BF_{\pi\pi}\equiv\BF_{\textnormal{\acs*{PDG}}}^{\textnormal{(fit)}}(\psi(2S)\to\pi^+\pi^- \Jpsi) = \qty{34.78(33)}{\percent}$~\cite{ParticleDataGroup:2026mpi} are taken from Ref.~\cite{CLEO:2008pln}.
The uncertainty of the line shape is estimated by varying the Breit--Wigner distribution in \cref{eq:line_shape} from a nonrelativistic to a relativistic one, resulting in a $\qty{7}{\percent}$ systematic uncertainty on $N_{\etac}$.
The remaining systematic uncertainties are taken over from Ref.~\cite{CLEO:2008pln}.

\subsection{\texorpdfstring{$\etac\to\gamma\gamma$}{etac -> gamma gamma} via \texorpdfstring{$\Jpsi\to\gamma\etac$}{J/psi -> gamma etac}}
\begin{figure}[h]
\centering
\includegraphics{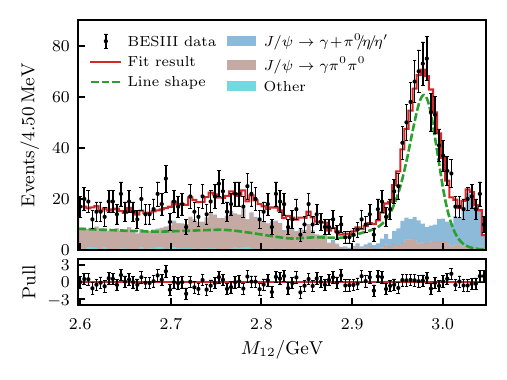}
\caption{%
Fit to the two-photon invariant mass ($M_{12}$) distribution from exclusive $\Jpsi\to\gamma\etac\to\gamma\gamma\gamma$ decays.
The black dots with error bars are data points measured by the BESIII experiment~\cite{BESIII:2024rex}.
The red solid curve shows the fit result, with the green dashed curve representing the line shape as described by \cref{eq:line_shape,eq:line_shape:observed}.
The blue, brown, and cyan (close to zero almost everywhere) filled histograms represent the simulated backgrounds from Ref.~\cite{BESIII:2024rex}.
The lower panel shows the pulls $(N_{\textnormal{data}}-N_{\textnormal{fit}})/\sqrt{\smash[b]{\sigma_{\textnormal{data}}^2 + \sigma_{\textnormal{tpl}}^2}}$, where $\sigma_{\textnormal{data}}=\sqrt{N_{\textnormal{data}}}$ denotes the statistical measurement uncertainty and $\sigma_{\textnormal{tpl}}$ the estimated template uncertainty.
}\label{fig:bes3:gg}
\end{figure}

We perform an extended binned maximum likelihood fit following the procedure outlined in Ref.~\cite{BESIII:2024rex} to the two-photon invariant mass ($M_{\gamma\gamma} \equiv M_{12}$) distribution based on the data points and background simulations presented in Fig.~2\,(d) of Ref.~\cite{Li:2025pri}.
The photon energy is related to the invariant mass via \cref{eq:momentum_conservation}, where $M_{\Jpsi}=\qty{3096.9}{\MeV}$~\cite{ParticleDataGroup:2026mpi}.
The photon energy spectrum line shape is described by \cref{eq:line_shape,eq:line_shape:observed}, and $D(M_{12} \given M)$ is modeled by a Gaussian resolution function with free parameters for the mass shift and resolution.
We take the mass-dependent efficiency $\epsilon(M)$ from Fig.~4\,(a) of Ref.~\cite{Li:2025pri}.
In analogy to Ref.~\cite{BESIII:2024rex}, we impose Gaussian constraints on the $\etac$ mass $M_{\etac}$ and width $\Gamma_{\etac}$ to the respective \ac{PDG} values~\cite{ParticleDataGroup:2026mpi}.
In addition, we impose Gaussian constraints on the background strengths to the total simulated yields with a \qty{10}{\percent} uncertainty.
However, since the new line shape is expected to capture the process $\Jpsi\to\gamma\gamma\gamma$ and potentially also processes referred to as \enquote{other minor contributions} in Ref.~\cite{BESIII:2024rex}, we relax their respective constraints by assigning a \qty{100}{\percent} uncertainty.
The best fit is shown in \cref{fig:bes3:gg}. 
The best-fit parameters are $\hfs=\qty{112.83(30)}{\MeV}$ and $\Gamma_{\etac}=\qty{30.16(50)}{\MeV}$.
The $\Jpsi\to\gamma\gamma\gamma$ background component is completely absorbed into the signal, which is consistent with the definition of the line shape, and the background component corresponding to other minor contributions is significantly reduced.

Weighting the mass-dependent efficiency with the fitted line shape, the signal efficiency is estimated to be $\epsilon_{\textnormal{sig}}=\qty{13.09(2)}{\percent}$, and using \cref{eq:signal_yield}, the signal yield is determined to be $N_{\etac}=\num{652.1(366)}$.
The product branching fraction is obtained from ($N_{\textnormal{sig}} \equiv N_{\etac}$)
\begin{equation}
	\BF_{\textnormal{BESIII}}^{\textnormal{(\acs*{M1},$\gamma\gamma$)}}
	=
	\frac{N_{\textnormal{sig}}}{N_{\psi(3686)}\epsilon_{\textnormal{sig}}\BF_{\pi\pi}}
	\,,
\end{equation}
where $N_{\psi(3686)}$ is taken from Ref.~\cite{BESIII:2024rex}.
As before, the uncertainty of the line shape is estimated by substituting the nonrelativistic Breit--Wigner distribution with a relativistic one, resulting in a \qty{9.2}{\percent} systematic uncertainty on $\BF_{\textnormal{BESIII}}^{\textnormal{(\acs*{M1},$\gamma\gamma$)}}$, whereas all remaining systematic uncertainties are adopted from Ref.~\cite{BESIII:2024rex}.

\subsection{\texorpdfstring{$\etac\to p\bar{p}$}{etac -> p pbar} via \texorpdfstring{$\Jpsi\to\gamma\etac$}{J/psi -> gamma etac}}
\begin{figure}[h]
\centering
\includegraphics{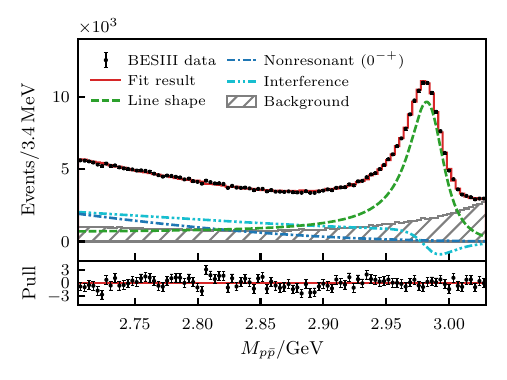}
\caption{%
Fit to the proton-antiproton invariant mass ($M_{p\bar{p}}$) distribution from exclusive $\Jpsi\to\gamma\etac\to\gamma p\bar{p}$ decays.
The black dots with error bars are data points measured by the BESIII experiment~\cite{BESIII:2025vdn}.
The red solid curve shows the fit result, with the green dashed curve representing the line shape as described by \cref{eq:line_shape,eq:line_shape:observed}, the blue dash-dotted curve representing the dominant nonresonant contribution, and the cyan dash-double-dotted curve representing their interference.
The gray hatched histogram represents the simulated background from Ref.~\cite{BESIII:2025vdn}.
The lower panel shows the pulls $(N_{\textnormal{data}}-N_{\textnormal{fit}})/\sqrt{\smash[b]{\sigma_{\textnormal{data}}^2 + \sigma_{\textnormal{tpl}}^2}}$, where $\sigma_{\textnormal{data}}=\sqrt{N_{\textnormal{data}}}$ denotes the statistical measurement uncertainty and $\sigma_{\textnormal{tpl}}$ the estimated template uncertainty.
}\label{fig:bes3:pp}
\end{figure}

We perform an extended binned maximum likelihood fit following the procedure outlined in Ref.~\cite{BESIII:2025vdn} to the proton-antiproton invariant mass ($M_{p\bar{p}}$) distribution based on the data points and background simulation presented in Fig.~1\,(a) of Ref.~\cite{BESIII:2025vdn}.
The mass projection is modeled by the modulus squared of a coherent sum of effective complex amplitudes corresponding to the contributing partial waves.
The photon energy is related to the invariant mass via \cref{eq:momentum_conservation}, where $M_{\Jpsi}=\qty{3096.9}{\MeV}$~\cite{ParticleDataGroup:2026mpi}.
To reproduce \cref{eq:line_shape}, the resonant amplitude is described by a linear function of $E_\gamma(M_{p\bar{p}})$ times a complex nonrelativistic Breit--Wigner amplitude.
The nonresonant amplitude for the wave with $J^{PC}=0^{-+}$ is modeled by a linear function of $E_\gamma(M_{p\bar{p}})$ times a Blatt--Weisskopf barrier factor described in Refs.~\cite{BESIII:2025vdn,Asner:2008nq}.
The respective intensities and the interference are multiplied by a phase space factor.
Detector resolution effects are described by $D(M_{p\bar{p}} \given M)$, which is modeled by a Gaussian resolution function with a mass shift and resolution of \qty{1.01}{\MeV} and \qty{3.93}{\MeV}, respectively~\cite{BESIII:2025vdn}.
The mass-dependent efficiency is assumed to be a constant below \qty{3.03}{\GeV}~\footnote{
	Y.~Liao and Y.~Zeng (private communication).
}.
A Gaussian constraint is imposed on the background strength to the total simulated yield with a \qty{10}{\percent} uncertainty.
The best fit is shown in \cref{fig:bes3:pp}. 
The best-fit parameters are $\hfs=\qty{109.23(11)}{\MeV}$ and $\Gamma_{\etac}=\qty{29.33(29)}{\MeV}$.
In principle, the contributions of other partial waves could introduce sizable uncertainties that cannot be determined in a one-dimensional fit.
However, the analysis in Ref.~\cite{BESIII:2025vdn} (albeit using a different line shape) suggests that they are minor compared to the $0^{-+}$ component.
A full amplitude analysis employing the proposed line shape would, of course, provide a more reliable description.

The observed signal yield is obtained using \cref{eq:signal_yield}, giving $N_{\etac}=\num{126.53(245)e3}$.
The product branching fraction is obtained from ($N_{\textnormal{sig}} \equiv N_{\etac}$)
\begin{equation}
	\BF_{\textnormal{BESIII}}^{\textnormal{(\acs*{M1},$p\bar{p}$)}}
	=
	\frac{N_{\textnormal{sig}}}{N_{\Jpsi}\epsilon_{\textnormal{sig}}}
	\,,
\end{equation}
where $N_{\Jpsi}$ and $\epsilon_{\textnormal{sig}}$ are taken from Ref.~\cite{BESIII:2025vdn}.
The systematic uncertainties are adopted from Ref.~\cite{BESIII:2025vdn}.

\section{Comparison}
\begin{figure}[h]
\centering
\includegraphics{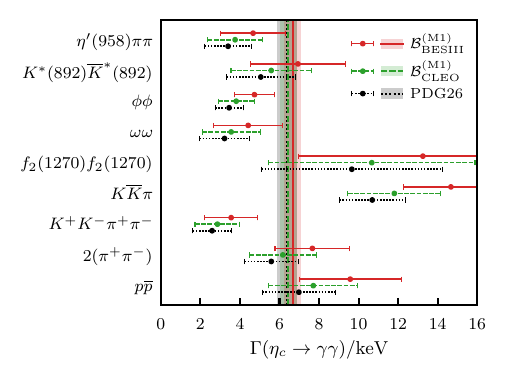}
\caption{%
Comparison of experimental values of the partial decay width $\Gamma(\etac\to\gamma\gamma)$ for different values of the branching fraction $\BF^{\textnormal{(M1)}}\equiv\BF(\Jpsi\to\gamma\etac)$.
For each $\etac$ decay channel $X_i$ listed on the left, the red/green/black dots with solid/dashed/dotted error bars show the determination of $[\Gamma(\etac\to X_i)\times\Gamma(\etac\to\gamma\gamma)/\Gamma_{\etac}]_{\textnormal{\acs*{PDG}}}^{\textnormal{(av)}}$ divided by $[\Gamma(\etac\to X_i)/\Gamma_{\etac}\times\Gamma(\Jpsi\to\gamma\etac)/\Gamma_{\Jpsi}]_{\textnormal{\acs*{PDG}}}^{\textnormal{(av)}}$ and multiplied by $\BF^{\textnormal{(M1)}}_{\textnormal{BESIII}}$ from \cref{eq:bfs:bes3:M1}, by $\BF^{\textnormal{(M1)}}_{\textnormal{CLEO}}$ from \cref{eq:bf:cleo}, or by $\BF_{\textnormal{\acs*{PDG}}}^{\text{(fit)}}(\Jpsi\to\gamma\etac)$~\cite{ParticleDataGroup:2026mpi}.
The vertical red-solid/green-dashed/black-dotted lines denote the respective multiparticle fit values $\Gamma^{\textnormal{(fit)}}(\etac\to\gamma\gamma)$, with the shaded bands indicating their uncertainties.
}\label{fig:comparison:gg}
\end{figure}

\begin{figure}[h]
\centering
\includegraphics{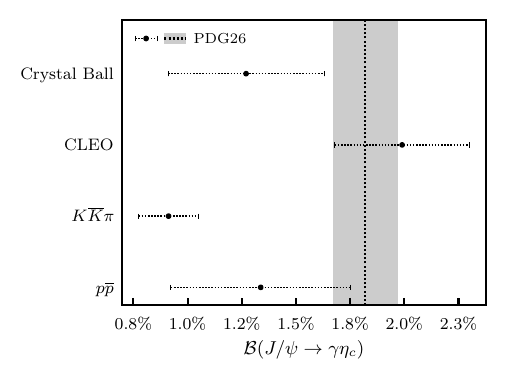}
\caption{%
Comparison of experimental values of the branching fraction $\BF(\Jpsi\to\gamma\etac)$.
The black dots with dotted error bars show either direct measurements of $\BF(\Jpsi\to\gamma\etac)$ (Crystal~Ball~\cite{Gaiser:1985ix}, CLEO~\cite{CLEO:2008pln}) or the determination of $[\Gamma(\etac\to X_i)/\Gamma_{\etac}\times\Gamma(\Jpsi\to\gamma\etac)/\Gamma_{\Jpsi}]_{\textnormal{\acs*{PDG}}}^{\textnormal{(av)}}$ divided by $[\Gamma(\etac\to X_i)/\Gamma_{\etac}]_{\textnormal{\acs*{PDG}}}^{\textnormal{(av)}}$, for the $\eta_c$ decay channel $X_i$ listed on the left.
The vertical black dotted line denotes the multiparticle fit value $\BF_{\textnormal{\acs*{PDG}}}^{\textnormal{(fit)}}(\Jpsi\to\gamma\etac)$~\cite{ParticleDataGroup:2026mpi}, with the shaded band indicating its uncertainty.
}\label{fig:comparison:m1}
\end{figure}

\Cref{fig:comparison:gg} shows the spread of experimental measurements that directly influence the fit value for $\Gamma(\etac\to\gamma\gamma)$ for various values of the branching fraction $\BF(\Jpsi\to\gamma\etac)$, 
and \cref{fig:comparison:m1} shows the spread of experimental measurements that directly influence the fit value for $\BF(\Jpsi\to\gamma\etac)$.
The relevant individual (product) branching fractions entering $[\Gamma(\etac\to X_i)/\Gamma_{\etac}\times\Gamma(\Jpsi\to\gamma\etac)/\Gamma_{\Jpsi}]_{\textnormal{\acs*{PDG}}}^{\textnormal{(av)}}$ and measurements of $\Gamma(\Jpsi\to\gamma\etac)/\Gamma_{\Jpsi}$ have been determined without using our prescription on the $\etac$ yield and, in some cases, even without using the correct photon energy spectrum line shape~\labelcref{eq:line_shape}, but just a Breit--Wigner distribution.
We suggest reanalyzing those processes with the method presented here.

\bibliographystyle{elsarticle-num}
\bibliography{letter}

@article{BESIII:2025vdn,
	Author = {Ablikim, Medina and others},
	Title = {{Study of the Magnetic Dipole Transition of $J/\psi\to\gamma\eta_c$ via $\eta_c\to p\bar{p}$}},
	Journal = {Phys. Rev. Lett.},
	Year = {2026},
	Volume = {136},
	Number = {5},
	Pages = {051901},
	DOI = {10.1103/r6yf-q6kv},
	EPRINT = {2510.15247},
	ARCHIVEPREFIX = {arXiv},
}

@article{BESIII:2026pff,
    Author = {Ablikim, Medina and others},
    Title = {{First Measurement of the Absolute Branching Fraction of $\eta_c\to\gamma\gamma$}},
    Journal = {Phys. Rev. Lett.},
    Year = {2026},
    Volume = {137},
    Number = {7},
    Pages = {071802},
    DOI = {10.1103/whyk-ykjw},
    EPRINT = {2601.11236},
    ARCHIVEPREFIX = {arXiv},
}

@article{Li:2025pri,
	Author = {Li, Zhijun and You, Zhengyun},
	Title = {{Observation of the decay $\eta_c\to\gamma\gamma$ at BESIII}},
	Journal = {PoS},
	Year = {2026},
	Volume = {EPS-HEP2025},
	Pages = {267},
	DOI = {10.22323/1.485.0267},
	EPRINT = {2510.26434},
	ARCHIVEPREFIX = {arXiv},
}

@article{ParticleDataGroup:2026mpi,
    Author = {Takahashi, F. and others},
    Title = {{Review of Particle Physics}},
    Journal = {Int. J. Mod. Phys. A},
    Year = {2026},
    Volume = {41},
    Number = {22},
    Pages = {2630011},
    DOI = {10.1142/S0217751X26300115},
}

@article{Beringer:2024eae,
	Author = {Beringer, Juerg},
	Title = {{New ways to access PDG data}},
	Journal = {PoS},
	Year = {2025},
	Volume = {ICHEP2024},
	Pages = {1023},
	DOI = {10.22323/1.476.1023},
}

@article{BESIII:2024rdn,
	Author = {Ablikim, Medina and others},
	Title = {{Observation of $\eta_c(1S,2S)$ and $\chi_{cJ}$ decays to $2(\pi^+\pi^-)\eta$ via $\psi(3686)$ radiative transitions}},
	Journal = {Phys. Rev. D},
	Year = {2025},
	Volume = {111},
	Number = {5},
	Pages = {052013},
	DOI = {10.1103/PhysRevD.111.052013},
	EPRINT = {2406.08225},
	ARCHIVEPREFIX = {arXiv},
}

@article{BESIII:2024rex,
	Author = {Ablikim, Medina and others},
	Title = {{Observation of the Charmonium Decay $\eta_c\to\gamma\gamma$ in $J/\psi\to\gamma\eta_c$}},
	Journal = {Phys. Rev. Lett.},
	Year = {2025},
	Volume = {134},
	Pages = {181901},
	DOI = {10.1103/PhysRevLett.134.181901},
	EPRINT = {2412.12998},
	ARCHIVEPREFIX = {arXiv},
}

@software{iminuit,
	Title = {scikit-hep/iminuit},
	Year = 2025,
	Author = {Dembinski, Hans and Ongmongkolkul, Piti and Deil, Christoph and Schreiner, Henry and Feickert, Matthew and Burr, Chris and Watson, Jason and Rost, Fabian and Pearce, Alex and Geiger, Lukas and Abdelmotteleb, Ahmed and Desai, Aman and Wiedemann, Bernhard M. and Gohlke, Christoph and Sanders, Jeremy and Drotleff, Jonas and Eschle, Jonas and Neste, Ludwig and Gorelli, Marco Edward and Baak, Max and Eliachevitch, Michael and Zapata, Omar},
	Month = 11,
	Url = {https://doi.org/10.5281/zenodo.17565861},
	Version = {v2.32.0},
	Publisher = {Zenodo},
	DOI = {10.5281/zenodo.17565861},
}

@article{Meng:2024axn,
	Author = {Meng, Yu and Liu, Chuan and Wang, Teng and Yan, Haobo},
	Title = {{Lattice study of $J/\psi\to\gamma\eta_c$ using a method without momentum extrapolation}},
	Journal = {Phys. Rev. D},
	Year = {2025},
	Volume = {111},
	Number = {1},
	Pages = {014508},
	DOI = {10.1103/PhysRevD.111.014508},
	EPRINT = {2411.04415},
	ARCHIVEPREFIX = {arXiv},
}

@article{Beringer:2024ady,
	Author = {Beringer, J.},
	Title = {{Programmatic access to PDG data}},
	Journal = {Nuovo Cim. C},
	Year = {2024},
	Number = {4},
	Pages = {206},
	DOI = {10.1393/ncc/i2024-24206-9},
}

@article{Delaney:2023fsc,
	Author = {Delaney, James and Thomas, Christopher E. and Ryan, Sinéad M.},
	Title = {{Radiative transitions in charmonium from lattice QCD}},
	Journal = {JHEP},
	Year = {2024},
	Volume = {05},
	Number = {2024},
	Pages = {230},
	DOI = {10.1007/JHEP05(2024)230},
	EPRINT = {2301.08213},
	ARCHIVEPREFIX = {arXiv},
}

@article{ParticleDataGroup:2024cfk,
	Author = {Navas, S. and others},
	Title = {{Review of Particle Physics}},
	Journal = {Phys. Rev. D},
	Year = {2024},
	Volume = {110},
	Number = {3},
	Pages = {030001},
	DOI = {10.1103/PhysRevD.110.030001},
}

@article{Colquhoun:2023zbc,
	Author = {Colquhoun, Brian and Cooper, Laurence J. and Davies, Christine T. H. and Lepage, G. Peter},
	Title = {{Precise determination of decay rates for $\eta_c \to \gamma\gamma$, $J/\psi\to\gamma\eta_c$, and $J/\psi\to\eta_ce^+e^-$ from lattice QCD}},
	Journal = {Phys. Rev. D},
	Year = {2023},
	Volume = {108},
	Number = {1},
	Pages = {014513},
	DOI = {10.1103/PhysRevD.108.014513},
	EPRINT = {2305.06231},
	ARCHIVEPREFIX = {arXiv},
}

@article{Meng:2021ecs,
	Author = {Meng, Yu and Feng, Xu and Liu, Chuan and Wang, Teng and Zou, Zuoheng},
	Title = {{First-principle calculation of the $\eta_c\to2\gamma$ decay width from lattice QCD}},
	Journal = {Sci. Bull.},
	Year = {2023},
	Volume = {68},
	Pages = {1880--1885},
	DOI = {10.1016/j.scib.2023.07.041},
	EPRINT = {2109.09381},
	ARCHIVEPREFIX = {arXiv},
}

@article{Segovia:2021bjb,
	Author = {Segovia, Jorge and Tarrús Castellà, Jaume},
	Title = {{Line shape and the experimental determination of the $J/\psi \to \gamma\eta_c$ branching fraction}},
	Journal = {Phys. Rev. D},
	Year = {2021},
	Volume = {104},
	Number = {7},
	Pages = {074032},
	DOI = {10.1103/PhysRevD.104.074032},
	EPRINT = {2106.15203},
	ARCHIVEPREFIX = {arXiv},
}

@article{CLQCD:2020njc,
	Author = {Chen, Ying and Gong, Ming and Li, Ning and Liu, Chuan and Liu, Yu-Bin and Liu, Zhaofeng and Ma, Jian-Ping and Meng, Yu and Xiong, Chao and Zhang, Ke-Long},
	Title = {{Lattice study of two-photon decay widths for scalar and pseudo-scalar charmonium}},
	Journal = {Chin. Phys. C},
	Year = {2020},
	Volume = {44},
	Number = {8},
	Pages = {083108},
	DOI = {10.1088/1674-1137/44/8/083108},
	EPRINT = {2003.09817},
	ARCHIVEPREFIX = {arXiv},
}

@article{Liu:2020qfz,
	Author = {Liu, Chuan and Meng, Yu and Zhang, Ke-Long},
	Title = {{Ward identity of the vector current and the decay rate of $\eta_c\rightarrow\gamma\gamma$ in lattice QCD}},
	Journal = {Phys. Rev. D},
	Year = {2020},
	Volume = {102},
	Number = {3},
	Pages = {034502},
	DOI = {10.1103/PhysRevD.102.034502},
	EPRINT = {2004.03907},
	ARCHIVEPREFIX = {arXiv},
}

@article{Gui:2019dtm,
	Author = {Gui, Long-Cheng and Dong, Jia-Mei and Chen, Ying and Yang, Yi-Bo},
	Title = {{Study of the pseudoscalar glueball in $J/\psi$ radiative decays}},
	Journal = {Phys. Rev. D},
	Year = {2019},
	Volume = {100},
	Number = {5},
	Pages = {054511},
	DOI = {10.1103/PhysRevD.100.054511},
	EPRINT = {1906.03666},
	ARCHIVEPREFIX = {arXiv},
}

@article{Anashin:2014wva,
	Author = {Anashin, V. V. and others},
	Title = {{Measurement of $J/\psi\to\gamma\eta_c$ decay rate and $\eta_c$ parameters at KEDR}},
	Journal = {Phys. Lett. B},
	Year = {2014},
	Volume = {738},
	Pages = {391--396},
	DOI = {10.1016/j.physletb.2014.09.064},
	EPRINT = {1406.7644},
	ARCHIVEPREFIX = {arXiv},
}

@article{Becirevic:2012dc,
	Author = {Be\v{c}irević, Damir and Sanfilippo, Francesco},
	Title = {{Lattice QCD study of the radiative decays $J/\psi\to \eta_c\gamma$ and $h_c\to \eta_c\gamma$}},
	Journal = {JHEP},
	Year = {2013},
	Volume = {01},
	Number = {2013},
	Pages = {028},
	DOI = {10.1007/JHEP01(2013)028},
	EPRINT = {1206.1445},
	ARCHIVEPREFIX = {arXiv},
}

@article{Pineda:2013lta,
	Author = {Pineda, Antonio and Segovia, J.},
	Title = {{Improved determination of heavy quarkonium magnetic dipole transitions in potential nonrelativistic QCD}},
	Journal = {Phys. Rev. D},
	Year = {2013},
	Volume = {87},
	Number = {7},
	Pages = {074024},
	DOI = {10.1103/PhysRevD.87.074024},
	EPRINT = {1302.3528},
	ARCHIVEPREFIX = {arXiv},
}

@article{Donald:2012ga,
	Author = {Donald, G. C. and Davies, C. T. H. and Dowdall, R. J. and Follana, E. and Hornbostel, K. and Koponen, J. and Lepage, G. P. and McNeile, C.},
	Title = {{Precision tests of the $J/{\psi}$ from full lattice QCD: Mass, leptonic width, and radiative decay rate to ${\eta}_c$}},
	Journal = {Phys. Rev. D},
	Year = {2012},
	Volume = {86},
	Pages = {094501},
	DOI = {10.1103/PhysRevD.86.094501},
	EPRINT = {1208.2855},
	ARCHIVEPREFIX = {arXiv},
}

@article{Brambilla:2010cs,
	Author = {Brambilla, N. and others},
	Title = {{Heavy quarkonium: progress, puzzles, and opportunities}},
	Journal = {Eur. Phys. J. C},
	Year = {2011},
	Volume = {71},
	Pages = {1534},
	DOI = {10.1140/epjc/s10052-010-1534-9},
	EPRINT = {1010.5827},
	ARCHIVEPREFIX = {arXiv},
}

@article{Brambilla:2010ey,
	Author = {Brambilla, Nora and Roig, Pablo and Vairo, Antonio},
	Title = {{Precise determination of the $\eta_c$ mass and width in the radiative $J/\psi \to \eta_c \gamma$ decay}},
	Journal = {AIP Conf. Proc.},
	Year = {2011},
	Volume = {1343},
	Pages = {418--420},
	DOI = {10.1063/1.3575048},
	EPRINT = {1012.0773},
	ARCHIVEPREFIX = {arXiv},
}

@article{Chen:2011kpa,
	Author = {Chen, Ying and others},
	Title = {{Radiative transitions in charmonium from $N_f=2$ twisted mass lattice QCD}},
	Journal = {Phys. Rev. D},
	Year = {2011},
	Volume = {84},
	Pages = {034503},
	DOI = {10.1103/PhysRevD.84.034503},
	EPRINT = {1104.2655},
	ARCHIVEPREFIX = {arXiv},
}

@article{Anashin:2010nr,
	Author = {Anashin, V. V. and others},
	Title = {{Measurement of $J/\psi \to \eta_c \gamma$ at KEDR}},
	Journal = {Chin. Phys. C},
	Year = {2010},
	Volume = {34},
	Pages = {831--835},
	DOI = {10.1088/1674-1137/34/6/035},
	EPRINT = {1002.2071},
	ARCHIVEPREFIX = {arXiv},
}

@article{Asner:2008nq,
	Author = {Asner, D. M. and others},
	Title = {{Physics at BES-III: Analysis Tools}},
	Journal = {Int. J. Mod. Phys. A},
	Year = {2009},
	Volume = {24},
	Number = {supp01},
	Pages = {23--77},
	DOI = {10.1142/S0217751X09046436},
	EPRINT = {0809.1869},
	ARCHIVEPREFIX = {arXiv},
}

@article{CLEO:2008pln,
	Author = {Mitchell, R. E. and others},
	Title = {{$J/\psi$ and $\psi(2S)$ Radiative Transitions to $\eta_c$}},
	Journal = {Phys. Rev. Lett.},
	Year = {2009},
	Volume = {102},
	Pages = {011801},
	DOI = {10.1103/PhysRevLett.102.011801},
	EPRINT = {0805.0252},
	ARCHIVEPREFIX = {arXiv},
}

@article{Dudek:2009kk,
	Author = {Dudek, Jozef J. and Edwards, Robert G. and Thomas, Christopher E.},
	Title = {{Exotic and excited-state radiative transitions in charmonium from lattice QCD}},
	Journal = {Phys. Rev. D},
	Year = {2009},
	Volume = {79},
	Pages = {094504},
	DOI = {10.1103/PhysRevD.79.094504},
	EPRINT = {0902.2241},
	ARCHIVEPREFIX = {arXiv},
}

@article{Brambilla:2005zw,
	Author = {Brambilla, Nora and Jia, Yu and Vairo, Antonio},
	Title = {{Model-independent study of magnetic dipole transitions in quarkonium}},
	Journal = {Phys. Rev. D},
	Year = {2006},
	Volume = {73},
	Pages = {054005},
	DOI = {10.1103/PhysRevD.73.054005},
	EPRINT = {hep-ph/0512369},
	ARCHIVEPREFIX = {arXiv},
}

@article{Dudek:2006ej,
	Author = {Dudek, Jozef J. and Edwards, Robert G. and Richards, David G.},
	Title = {{Radiative transitions in charmonium from lattice QCD}},
	Journal = {Phys. Rev. D},
	Year = {2006},
	Volume = {73},
	Pages = {074507},
	DOI = {10.1103/PhysRevD.73.074507},
	EPRINT = {hep-ph/0601137},
	ARCHIVEPREFIX = {arXiv},
}

@techreport{QuarkoniumWorkingGroup:2004kpm,
	Author = {Brambilla, N. and others},
	Title = {{Heavy Quarkonium Physics}},
	Institution = {CERN},
	Year = {2005},
	Type = {CERN Yellow Report},
	Month = {12},
	DOI = {10.5170/CERN-2005-005},
	EPRINT = {hep-ph/0412158},
	ARCHIVEPREFIX = {arXiv},
}

@article{Manohar:2003xv,
	Author = {Manohar, Aneesh V. and Ruiz-Femenía, Pedro},
	Title = {{Orthopositronium decay spectrum using NRQED}},
	Journal = {Phys. Rev. D},
	Year = {2004},
	Volume = {69},
	Pages = {053003},
	DOI = {10.1103/PhysRevD.69.053003},
	EPRINT = {hep-ph/0311002},
	ARCHIVEPREFIX = {arXiv},
}

@article{Gaiser:1985ix,
	Author = {Gaiser, J. E. and others},
	Title = {{Charmonium spectroscopy from inclusive $\psi^\prime$ and $J/\psi$ radiative decays}},
	Journal = {Phys. Rev. D},
	Year = {1986},
	Volume = {34},
	Pages = {711},
	DOI = {10.1103/PhysRevD.34.711},
}

@article{James:1975dr,
	Author = {James, F. and Roos, M.},
	Title = {{Minuit -- a system for function minimization and analysis of the parameter errors and correlations}},
	Journal = {Comput. Phys. Commun.},
	Year = {1975},
	Volume = {10},
	Pages = {343--367},
	DOI = {10.1016/0010-4655(75)90039-9},
}

@article{E598:1974sol,
	Author = {Aubert, J. J. and others},
	Title = {{Experimental Observation of a Heavy Particle $J$}},
	Journal = {Phys. Rev. Lett.},
	Year = {1974},
	Volume = {33},
	Pages = {1404--1406},
	DOI = {10.1103/PhysRevLett.33.1404},
}

@article{SLAC-SP-017:1974ind,
	Author = {Augustin, J.-E. and others},
	Title = {{Discovery of a Narrow Resonance in $e^+ e^-$ Annihilation}},
	Journal = {Phys. Rev. Lett.},
	Year = {1974},
	Volume = {33},
	Pages = {1406--1408},
	DOI = {10.1103/PhysRevLett.33.1406},
}

\end{document}